\documentclass[a4paper]{article}

\title{Three-phase point in a binary hard-core lattice model?}
\author{Alain Verberkmoes and Bernard Nienhuis\\
  \sl Instituut voor Theoretische Fysica, Universiteit van Amsterdam,\\
  \sl Valckenierstraat 65, 1018 XE Amsterdam, The Netherlands}
\date{March 25, 1999}

\usepackage{a4wide}
\usepackage{graphicx}

\begin{document}

%%%%%%%%%%%%%%%%%%%%%%%%%%%%%%%%%%%%%%%%%%%%%%%%%%%%%%%%%%%%%%%%%%%%%%%%

\maketitle

\begin{abstract}
\noindent
Using Monte Carlo simulation, Van Duijneveldt and Lekkerkerker [Phys.\
Rev.\ Lett.\ {\bf 71}, 4264 (1993)] found gas--liquid--solid behaviour
in a simple two-dimensional lattice model with two types of hard
particles.  The same model is studied here by means of numerical
transfer matrix calculations, focusing on the finite size scaling of
the gaps between the largest few eigenvalues.  No evidence for a
gas--liquid transition is found.  We discuss the relation of the model
with a solvable RSOS model of which the states obey the same exclusion
rules.  Finally, a detailed analysis of the relation with the dilute
three-state Potts model strongly supports the tricritical point rather
than a three-phase point.
\end{abstract}

%%%%%%%%%%%%%%%%%%%%%%%%%%%%%%%%%%%%%%%%%%%%%%%%%%%%%%%%%%%%%%%%%%%%%%%%

\section{Introduction}

The phase behaviour of hard particles, in particular spheres, as a
simple model of interacting particles, has received much attention.
Computer simulations of monodisperse hard spheres show a first-order
transition between a dilute disordered phase (fluid) and a dense
ordered phase (solid)~\cite{wood:1957,alder:1957,hoover:1968}.  The
continuous translational symmetry of the Hamiltonian remains intact in
the fluid, but is broken to a discrete subgroup in the solid.  Although
a rigorous proof is lacking, this phase transition in the hard sphere
model is now generally accepted.  For bidisperse hard spheres the
situation is more complicated.  The existence of several solid phases
has been established; see for example~\cite{eldridge:1993} and the
references therein.  The behaviour in the fluid phase, however, is not
known.  Using the Percus--Yevick closure of the Ornstein--Zernike
equation, Lebowitz and Rowlinson~\cite{lebowitz:1964} found miscibility
in all proportions for all diameter ratios.  More recently however,
Biben and Hansen~\cite{biben:1991}, using the Rogers--Young closure,
found a spinodal instability when the diameter ratio exceeds~5.  Even
so it might be that the fluid--fluid transition is pre-empted by the
fluid--solid transition, so that the former does not actually occur.
Thus it remains an open question whether bidisperse spheres can show a
fluid--fluid phase separation.  More generally one may ask if
gas--liquid--solid behaviour can occur in binary mixtures with only
hard-core repulsion.

\begin{figure}
  \hfil\includegraphics[width=0.8 \textwidth]{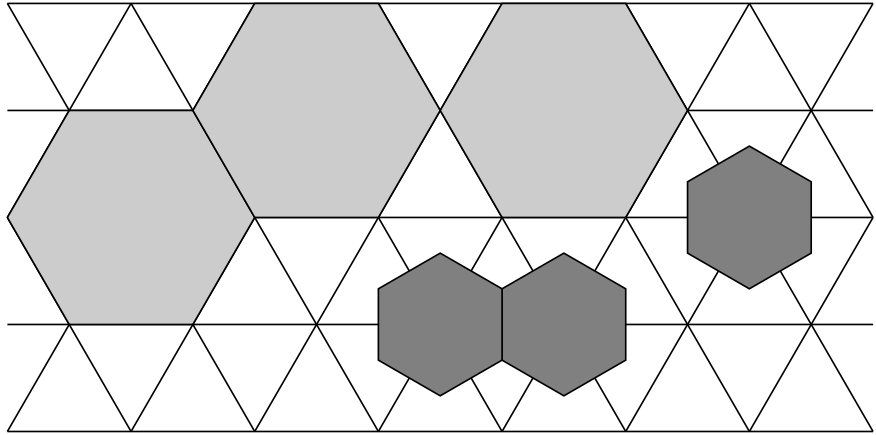}\hfil
  \caption{A typical configuration of large and small hexagons.}
  \label{fig:hexagons:model}
\end{figure}

\begin{figure}
  \hfil\includegraphics[width=0.8 \textwidth]{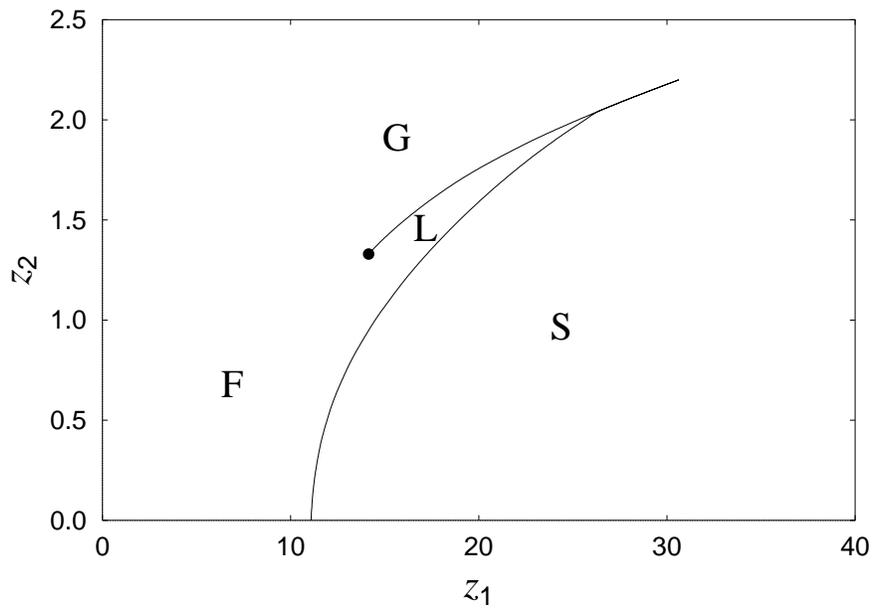}\hfil
  \caption{Phase diagram in the $z_1$--$z_2$ plane calculated by Van
    Duijneveldt and Lekkerkerker~\protect\cite{duijneveldt:1993,%
    duijneveldt:1995} from Monte Carlo simulations.  The letters F, G,
    L and S indicate the fluid, gas, liquid and solid phase,
    respectively.}
  \label{fig:hexagons:duin}
\end{figure}

Motivated by this interest Van Duijneveldt and
Lekkerkerker~\cite{duijneveldt:1993,duijneveldt:1995} studied a
two-dimensional binary hard-core lattice model.  This model, introduced
by Frenkel and Louis~\cite{frenkel:1992}, consists of large and small
hard hexagons on a triangular lattice, see
Figure~\ref{fig:hexagons:model}.  Every site can be empty or occupied
by a large or small hexagon, and if it is occupied with a large hexagon
all its direct neighbours must be empty.  When the small particles are
omitted, one regains the hard hexagon model~\cite{burley:1960}, which
has been solved exactly by Baxter~\cite{baxter:1980,baxter:1982}; it
has a second-order ordering transition.  Van Duijneveldt and
Lekkerkerker studied the binary model by means of Monte~Carlo
simulation.  They found three phases: dilute disordered (gas), dense
disordered (liquid), and ordered (solid).
Figure~\ref{fig:hexagons:duin} shows this phase diagram, represented in
terms of the fugacities $z_1$ and $z_2$ of the large and small
hexagons, respectively.

In this paper we study the same model by different methods.  Our
interest is in the qualitative, rather than quantitative, aspects of
the phase diagram.  We do not address the general question whether
gas--liquid--solid behaviour is possible in binary hard-core mixtures.
The paper is organised as follows.  First we briefly review the Monte
Carlo approach of Van Duijneveldt and Lekkerkerker, and we give some
exact results.  Then we describe our numerical transfer matrix
calculations.  Next we discuss the relation of the model with an
exactly solvable RSOS model and with the dilute three-state Potts
model.  Finally we propose an explanation for the discrepancy between
our results and those of Van Duijneveldt and Lekkerkerker.

%%%%%%%%%%%%%%%%%%%%%%%%%%%%%%%%%%%%%%%%%%%%%%%%%%%%%%%%%%%%%%%%%%%%%%%%

\section{Monte Carlo simulation and exact results}

Before we review the Monte~Carlo method of Van Duijneveldt and
Lekkerkerker~\cite{duijneveldt:1993,duijneveldt:1995} and discuss some
exact results, we make the following notational conventions: the
subscripts 1 and 2 refer to the large and small hexagons, respectively;
the superscript 0 refers to the pure hard hexagon model; the symbol $N$
without subscript is the number of sites and is generally omitted as an
argument of the thermodynamic quantities.

We consider the semi-grand canonical partition function $Z(N_1,z_2 )$
of large hexagons, whose number $N_1$ is fixed, and small hexagons,
whose fugacity $z_2$ is fixed, on $N$ lattice sites.  We may view the
small hexagons as causing an effective so-called depletion
interaction~\cite{asakura:1954}
between the large hexagons.  The question is then if this attractive
depletion interaction is strong enough to induce a fluid--fluid
transition.  The effective interaction can be expressed in the number
of sites available for small hexagons, once the large hexagons have
been placed on the lattice.  Interestingly, the sites available for
small hexagons are exactly the sites where an additional large hexagon
could be inserted.  Such sites are called free.  It is easy to express
the semi-grand canonical partition function $Z(N_1,z_2 )$ in terms of
the canonical partition function $Z^0(N_1 )$ of the hard hexagon model
and the probability distribution $p(N_{\mathrm{f}}|N_1 )$ for the
number $N_{\mathrm{f}}$ of free lattice sites in the hard hexagon
model:
\begin{displaymath}
  Z(N_1,z_2 ) = Z^0(N_1 ) Z'(N_1,z_2 ),
\end{displaymath}
where
\begin{displaymath}
  Z'(N_1,z_2 ) = \sum_{N_{\mathrm{f}}} \, p(N_{\mathrm{f}}|N_1 )
  (1+z_2)^{N_{\mathrm{f}}}.
\end{displaymath}
After taking logarithms this gives the free energy:
\begin{equation}
  F(N_1,z_2 ) = F^0(N_1 ) + F'(N_1,z_2 ),
  \label{equ:hexagons:freeen}
\end{equation}
Van Duijneveldt and Lekkerkerker determine the probability distribution
$p$ from canonical Monte Carlo simulations of the hard hexagon model.
To determine accurately the wings of the distribution an umbrella
sampling technique is employed.  They calculate $F'$ from $p$, and for
fixed $z_2$ fit a polynomial in $\rho_1:\,=N_1/N$ to this quantity.
They obtain the free energy $F$ from (\ref{equ:hexagons:freeen}), using
Baxter's exact result~\cite{baxter:1980,baxter:1982} for $F^0$ and the
fitted polynomial for~$F'$.  The fugacity $z_1$ of the large hexagons
and the pressure $P$ are calculated in the usual way from~$F$.  Finally
phase equilibrium is determined by looking for phases with equal $z_1$
and $P$ but different~$\rho_1$.  As this calculation is carried out for
fixed~$z_2$, $z_2$ is also equal in the phases.  The resulting phase
diagram is shown in Figure~\ref{fig:hexagons:duin}.  It has three
branches: liquid--solid, gas--solid and gas--liquid.  The branches meet
at the three-phase point, at $z_1=22.5$ and $z_2=1.89$.  (Van
Duijneveldt and Lekkerkerker use the term ``triple point'', but as that
suggests the coexistence of three phases where three first-order
transitions meet we prefer to use the term ``three-phase point''.)  The
gas--liquid end-point is located at $z_1=13.3$ and $z_2=1.36$.

Expanding $Z'$ to first order in $z_2$ gives:
\begin{equation}
  Z'(N_1,z_2 ) = 1 + z_2 \langle N_{\mathrm{f}} \rangle^0_{N_1 }
  + o(z_2).
  \label{equ:hexagons:exa}
\end{equation}
For a finite system we could have written ${\cal O}(z_2^2)$ instead of
$o(z_2)$, but in the thermodynamic limit this is not valid at the phase
transition of the hard hexagon model.  Lekkerkerker (unpublished) found
that the average $ \rho_{\mathrm{f}}:\,=\langle N_{\mathrm{f}}/N
\rangle^0_{N_1 }$ can be calculated exactly, as follows.  Adding one
hexagon to a configuration of $N_1$ hexagons can be done in
$N_{\mathrm{f}}$ ways.  By doing this to all configurations of $N_1$
hexagons each configuration of $N_1+1$ hexagons is obtained exactly
$N_1+1$ times.  Hence
\begin{displaymath}
  \langle N_{\mathrm{f}} \rangle^0_{N_1 } Z^0(N_1 ) =
  (N_1+1) Z^0(N_1+1 ),
\end{displaymath}
which in the thermodynamic limit yields
\begin{equation}
  \rho_{\mathrm{f}} = {\rho_1 \over z_1}.
  \label{equ:hexagons:exb}
\end{equation}
This is an example of Widom's famous particle-insertion
formula~\cite{widom:1963}.  In the Appendix we apply this exact result
in the method of Van Duijneveldt and Lekkerkerker.  In particular we
show that the existence of a Van der Waals loop cannot be concluded
from its presence in the first order
approximant~(\ref{equ:hexagons:exa}).

As the first derivatives of the thermodynamic functions with respect to
$z_2$ are known in this way, we shall now attempt to calculate the
locus of the phase transition in this order.  The difference between
the large and small hexagons is, that two small hexagons may occupy
neighbouring sites, whereas two large ones may not.  At small $z_2$ the
density of small hexagons is low, so that they will generally occur
isolated.  Thus they cannot be distinguished from the large ones.  For
the grand canonical partition function this implies:
\begin{equation}
  Z(z_1,z_2 ) = Z^0(z_1+z_2 ) + o(z_2).
  \label{equ:hexagons:pfun}
\end{equation}
This suggests that the locus of the phase transition is given by
\begin{equation}
  z_1 = z_1^{\mathrm{c}} - z_2 + o(z_2),
  \label{equ:hexagons:locz}
\end{equation}
where the superscript $\mathrm{c}$ refers to the critical point of the
pure hard hexagon model.  The particle densities follow also:
\begin{displaymath}
  \rho_1(z_1,z_2) = {z_1 \over z_1+z_2} \rho_1^0(z_1+z_2) + o(z_2)
\end{displaymath}
for the large hexagons, and similarly for the small ones.  Combining
these results yields the density of the large hexagons at the phase
transition:
\begin{equation}
  \rho_1 = \left( 1 - {z_2 \over z_1^{\mathrm{c}}} \right)
  \rho_1^{\mathrm{c}} + o(z_2).
  \label{equ:hexagons:locrho}
\end{equation}
Equations (\ref{equ:hexagons:locz}) and (\ref{equ:hexagons:locrho})
cannot be derived rigorously from (\ref{equ:hexagons:pfun}) alone, but
we conjecture that they are nevertheless valid.

%%%%%%%%%%%%%%%%%%%%%%%%%%%%%%%%%%%%%%%%%%%%%%%%%%%%%%%%%%%%%%%%%%%%%%%%

\section{Transfer matrix approach}
\label{sec:hexagons:transfer}

\begin{figure}
  \hfil\includegraphics[width=0.8 \textwidth]{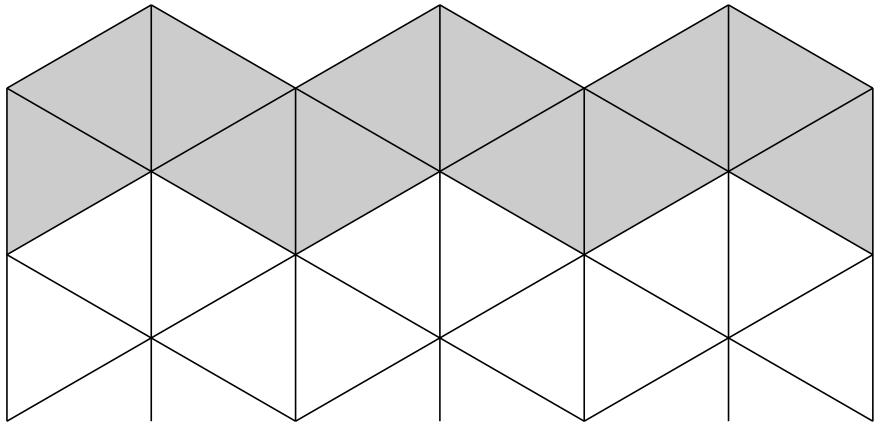}\hfil
  \caption{The transfer matrix adds one row (shaded) to the system.}
  \label{fig:hexagons:xfer}
\end{figure}

Now we study the model through its row-to-row transfer matrix.  For
practical reasons, we work with sawtooth rows as shown in
Figure~\ref{fig:hexagons:xfer}.  One advantage is that the high density
ground state of the hexagons fits on the lattice (which has an even
number of sites), whereas for straight rows it does so only when the
system size is a multiple of three.  Another advantage is that the
transfer matrix can be built up by repeatedly adding one site, without
increasing the total number of sites.  Periodic boundary conditions are
imposed on the rows.  The number of ``teeth'' is denoted by $W$, so a
row contains $2W$ sites and has length $L=W\sqrt{3}$.  The largest few
eigenvalues of the transfer matrix (in the zero-momentum sector) were
calculated numerically for $W=2$, \dots,~$5$, using the power method.

In the ordered regime there are in fact three coexisting ordered
phases, corresponding to the three sub-lattices of the triangular
lattice.  They give rise to three eigenvectors of the transfer matrix,
dominated by these ordered phases: one symmetric and two asymmetric
for permutations among the ground states.  The symmetric vector has the
largest eigenvalue~$\Lambda_0$.  The asymmetric vectors have a complex
conjugate pair of eigenvalues $\Lambda_{\mathrm{M}}$
and~$\Lambda_{\mathrm{M}}^*$.  In the region of the phase diagram we
are interested in, $\Lambda_0$, $\Lambda_{\mathrm{M}}$ and
$\Lambda_{\mathrm{M}}^*$, together with another real
eigenvalue~$\Lambda_{\mathrm{T}}$, turn out to be the largest
eigenvalues of the transfer matrix.  The phase behaviour can be
diagnosed from the behaviour of the gaps between the eigenvalues,
$\Delta_{\mathrm{M}}:\,=\log|\Lambda_0/\Lambda_{\mathrm{M}}|$ and
$\Delta_{\mathrm{T}}:\,=\log|\Lambda_0/\Lambda_{\mathrm{T}}|$, as the
system size $L$ tends to infinity.

The gap $\Delta_{\mathrm{T}}$ is an inverse correlation length between
density fluctuations.  In the absence of a phase transition, the bulk
($L=\infty$) value of this length is finite and the value for finite
$L$ approaches this bulk value when $L$ tends to infinity.  Hence
$\Delta_{\mathrm{T}}$ tends to a non-zero limit.  At a critical point
the bulk correlation length diverges and the value for finite $L$ is
proportional to~$L$.  As a consequence of scale invariance
$\Delta_{\mathrm{T}}$ decreases as~$1/L$.  At a first-order transition
with a change in the density, however, $\Delta_{\mathrm{T}}$ is not an
inverse correlation length.  The eigenvalues $\Lambda_0$ and
$\Lambda_{\mathrm{T}}$ are then asymptotically degenerate.  Their gap
$\Delta_{\mathrm{T}}$ is related to the interfacial tension between the
coexisting phases.  More precisely, $\Delta_{\mathrm{T}}$ decays as
$\exp(-\sigma L)$, where $\sigma$ is proportional to the interfacial
tension~\cite{fisher:1969}.

For the gap~$\Delta_{\mathrm{M}}$ the situation is analogous.  In the
disordered regime, it is an inverse correlation length, here between
fluctuations in the sub-lattice ordering.  Thus the gap approaches a
non-zero value as $L$ grows.  At a first-order transition between two
disordered phases this correlation length is generally different in the
two phases.  Therefore the value of $\Delta_{\mathrm{M}}$ undergoes a
sharp change through the transition, approaching a jump as the system
size $L$ increases.  At a critical point the bulk correlation length
diverges, so that $\Delta_{\mathrm{M}}$ decays as $1/L$ when $L$
increases.  In the ordered regime three phases coexist, and the
eigenvalues $\Lambda_0$ and $\Lambda_{\mathrm{M}}$ (and
$\Lambda_{\mathrm{M}}^*$) are asymptotically degenerate:
$\Delta_{\mathrm{M}}$ decays exponentially with~$L$.  At a first-order
transition between an ordered and a disordered phase by the same token
$\Delta_{\mathrm{M}}$ vanishes exponentially with~$L$.

\begin{figure}[!htb]
  \hfil\includegraphics[width=0.8 \textwidth]{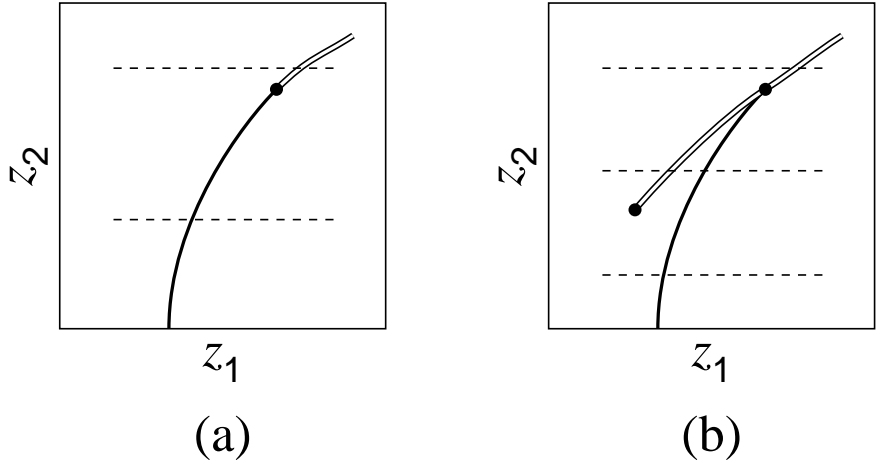}\hfil
  \caption{(a) Phase diagram with a fluid and solid phase.  The
    critical line (fat) terminates at a tricritical point where the
    phase transition becomes first-order (double line).  (b) Phase
    diagram with gas, liquid, and solid phases.  The critical line
    (fat) meets the first-order transition (double line) at the
    three-phase point.}
  \label{fig:hexagons:phases}
\end{figure}

We shall now distinguish between two scenarios: (i) there are two
phases (fluid and solid) as in Figure~\ref{fig:hexagons:phases}(a);
(ii) there are three phases (gas, liquid and solid) as in
Figure~\ref{fig:hexagons:phases}(b).  The gaps should behave as
follows.  At fixed $z_2$, the gap $\Delta_{\mathrm{M}}$ decreases with
increasing~$z_1$, whereas $\Delta_{\mathrm{T}}$ has a minimum at the
phase transition(s).  For low $z_2$, see the lower dashed lines in
Figures \ref{fig:hexagons:phases}(a) and \ref{fig:hexagons:phases}(b),
the scaled gaps $L\Delta_{\mathrm{M}}$ and $L\Delta_{\mathrm{T}}$ will
tend to a non-zero value when $L\to\infty$ at the critical line.  For
high $z_2$, see the upper dashed lines, this is no longer the case:
both scaled gaps tend to zero when $L\to\infty$ at the phase
transition, which is now first-order.  On the middle dashed line in
Figure~\ref{fig:hexagons:phases}(b), $\Delta_{\mathrm{M}}$ changes
rapidly at the gas-liquid transition.  Furthermore
$\Delta_{\mathrm{T}}$ has two minima: at the gas-liquid transition and
at the liquid-solid transition.  When $L\to\infty$, the minimum of the
scaled gap $L\Delta_{\mathrm{T}}$ tends to zero at the gas--liquid
transition, but to a non-zero value at the liquid--solid transition.
Thus the gas-liquid transition in Figure~\ref{fig:hexagons:phases}(b)
can be recognised from the appearance of a sudden change in
$\Delta_{\mathrm{M}}$ and a second minimum of~$\Delta_{\mathrm{T}}$.

\begin{figure}[p]
  \hfil\includegraphics[width=0.8 \textwidth]{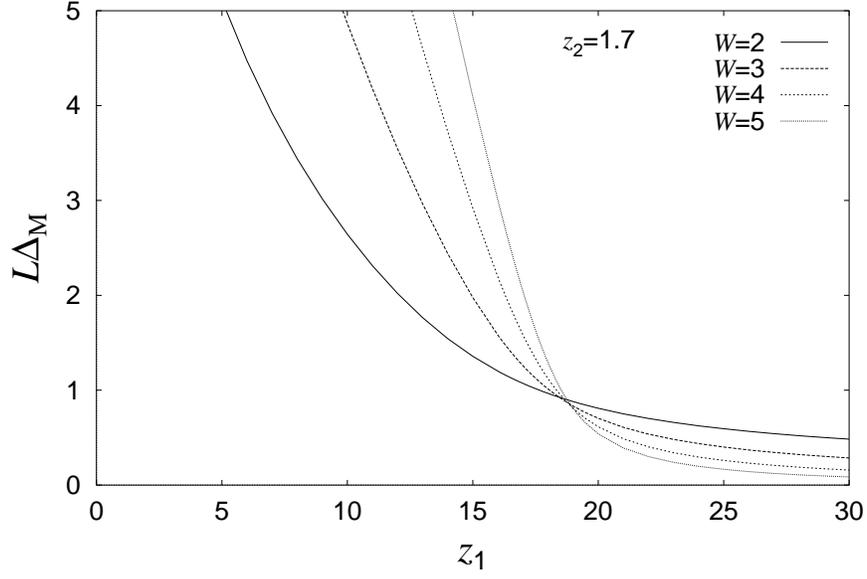}\hfil
  \caption{The scaled gaps $L\Delta_{\mathrm{M}}$ as a function of
    $z_1$ on the line~$z_2=1.7$.}
  \label{fig:hexagons:mlo}
\end{figure}

\begin{figure}[p]
  \hfil\includegraphics[width=0.8 \textwidth]{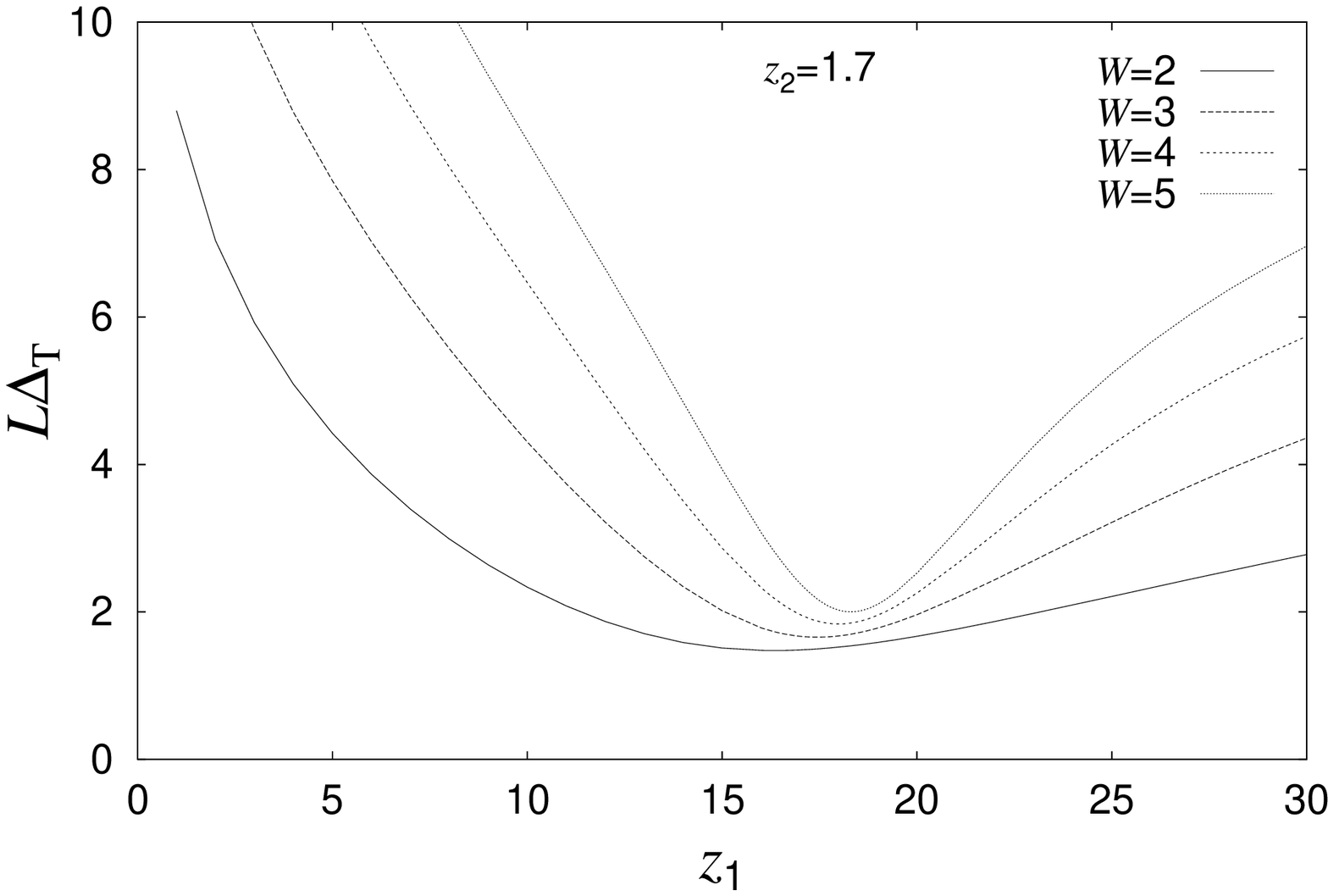}\hfil
  \caption{The scaled gaps $L\Delta_{\mathrm{T}}$ as a function of
    $z_1$ on the line~$z_2=1.7$.}
  \label{fig:hexagons:tlo}
\end{figure}

\begin{figure}[p]
  \hfil\includegraphics[width=0.8 \textwidth]{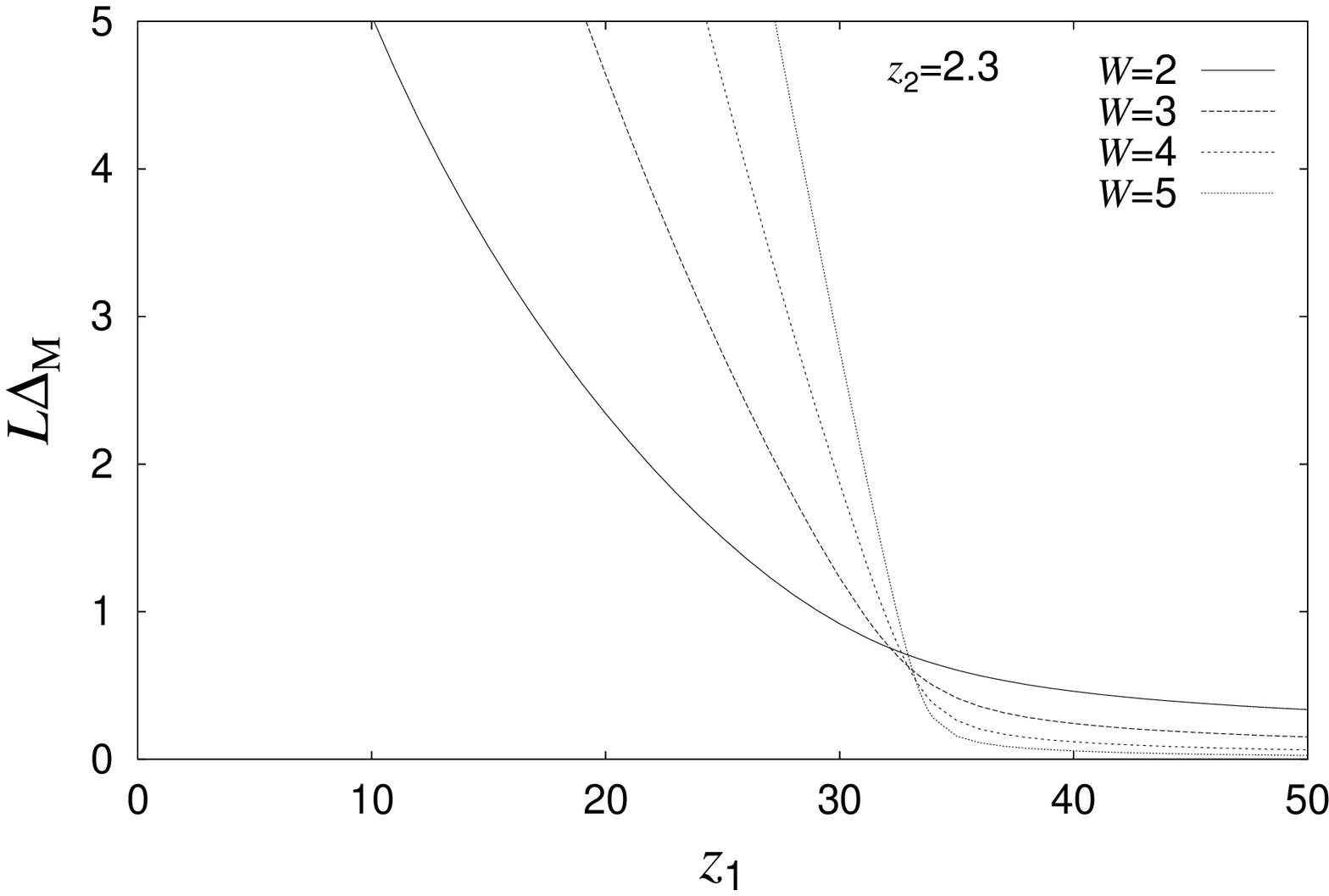}\hfil
  \caption{The scaled gaps $L\Delta_{\mathrm{M}}$ as a function of
    $z_1$ on the line~$z_2=2.3$.}
  \label{fig:hexagons:mhi}
\end{figure}

\begin{figure}[p]
  \hfil\includegraphics[width=0.8 \textwidth]{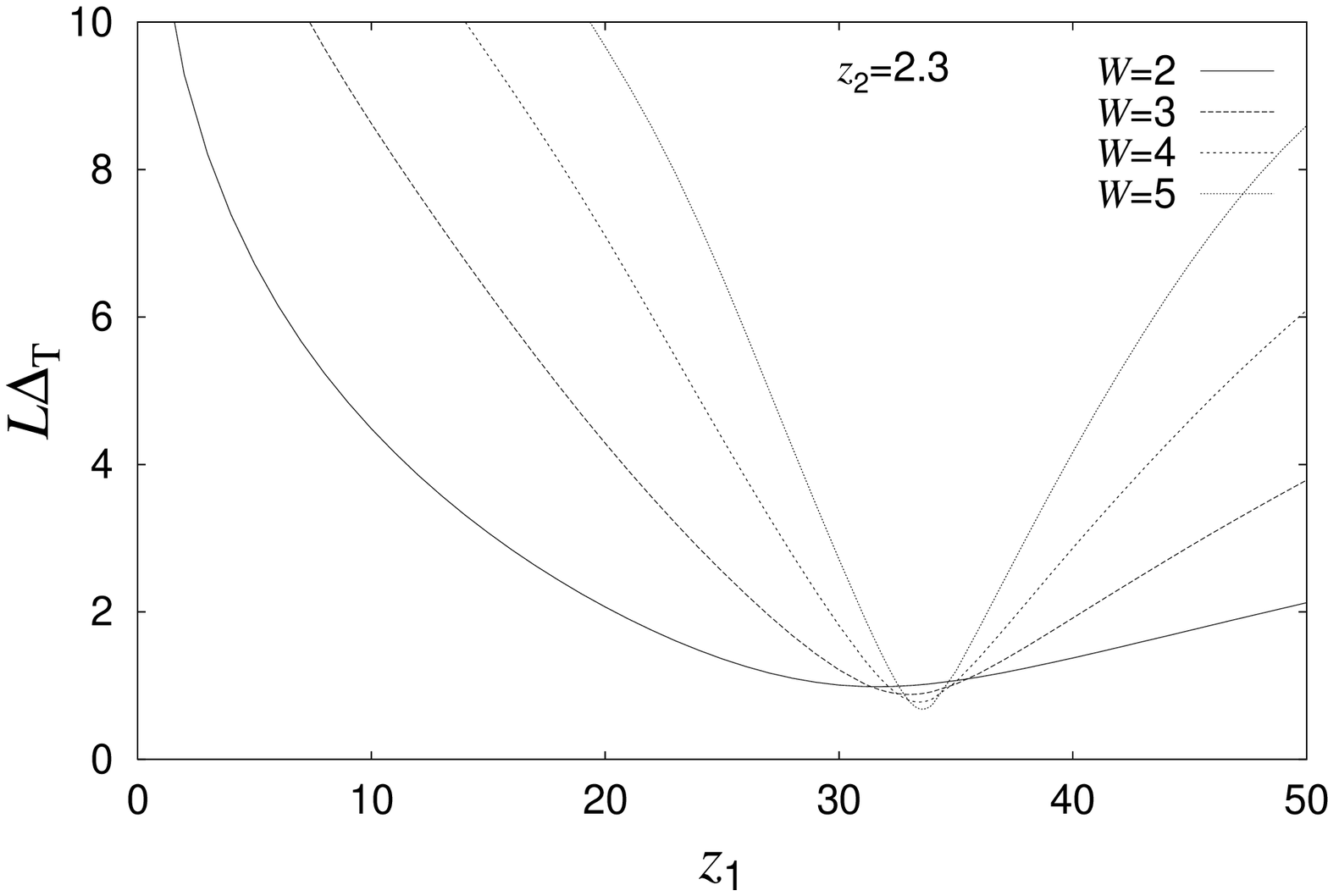}\hfil
  \caption{The scaled gaps $L\Delta_{\mathrm{T}}$ as a function of
    $z_1$ on the line~$z_2=2.3$.}
  \label{fig:hexagons:thi}
\end{figure}

For $z_2=0.0$, $0.1$, \dots,~$3.0$ the scaled gaps
$L\Delta_{\mathrm{M}}$ and $L\Delta_{\mathrm{T}}$ were plotted as
function of~$z_1$ for $W=2$ ,\dots,~$5$.  Figures
\ref{fig:hexagons:mlo}--\ref{fig:hexagons:thi}
show examples of this.  We found no indication that
$\Delta_{\mathrm{T}}$ has two minima.  One could argue that two minima
might be fused to a single one for these relatively small systems;
however, the sharpest and deepest minimum (at the gas--liquid
transition) is clearly absent.  This pleads against the three-phase
scenario in favour of the two-phase scenario.  We also saw no sudden
change in~$\Delta_{\mathrm{M}}$.  However, even if a gas--liquid
transition were present, the signal in $\Delta_{\mathrm{M}}$ might be
hard to detect.

\begin{figure}
  \hfil\includegraphics[width=0.8 \textwidth]{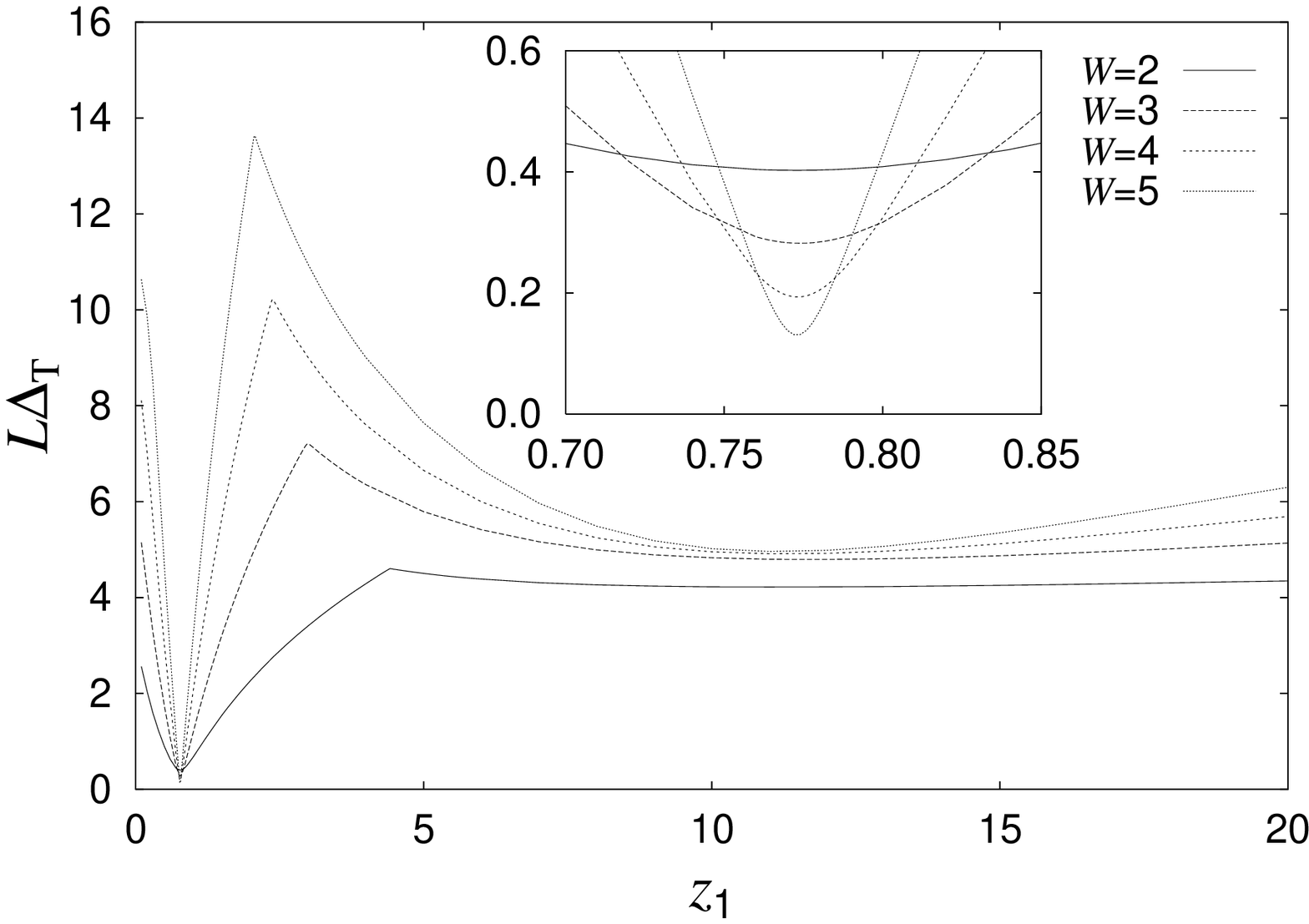}\hfil
  \caption{The scaled gaps $L\Delta_{\mathrm{T}}$ as a function of
    $z_1$ on the line~$z_2=1.3$ in the model with extra
    parameter~$\kappa=0.6$.  The inset shows the deep minima in more
    detail.}
  \label{fig:hexagons:tkappa}
\end{figure}

The three-phase scenario can be obtained by introducing an extra
parameter into the model.  Assign a weight $\kappa$ to every lattice
edge joining a small hexagon and an empty site.  For $\kappa=1$ one
recovers the original model.  For $\kappa=0$ any contact between a
small particle and an empty site is forbidden.  In this limit the model
either contains no small hexagons at all or is completely filled with
them.  The regime without small hexagons still exhibits the hard
hexagon transition as long as $1+z_2$ is smaller than the partition sum
per site of the hard hexagon model.  Beyond this value the phase filled
with small particles takes over.  Thus the ordered and disordered hard
hexagon phases meet with the pure small hexagon phase, where the phase
transition between them terminates in a three-phase point.  For
$\kappa$ close to zero the model will still obey the three-phase
scenario.  Here $\Delta_{\mathrm{T}}$ is indeed found to have two
minima, see Figure~\ref{fig:hexagons:tkappa}.  (The maxima in this
figure at first sight seem to be crossings of eigenvalues, but a very
close look reveals that they are in fact rounded.)  This supports our
interpretation of the absence of a second minimum in
$\Delta_{\mathrm{T}}$ as evidence against the three-phase scenario.

\begin{figure}
  \hfil\includegraphics[width=0.8 \textwidth]{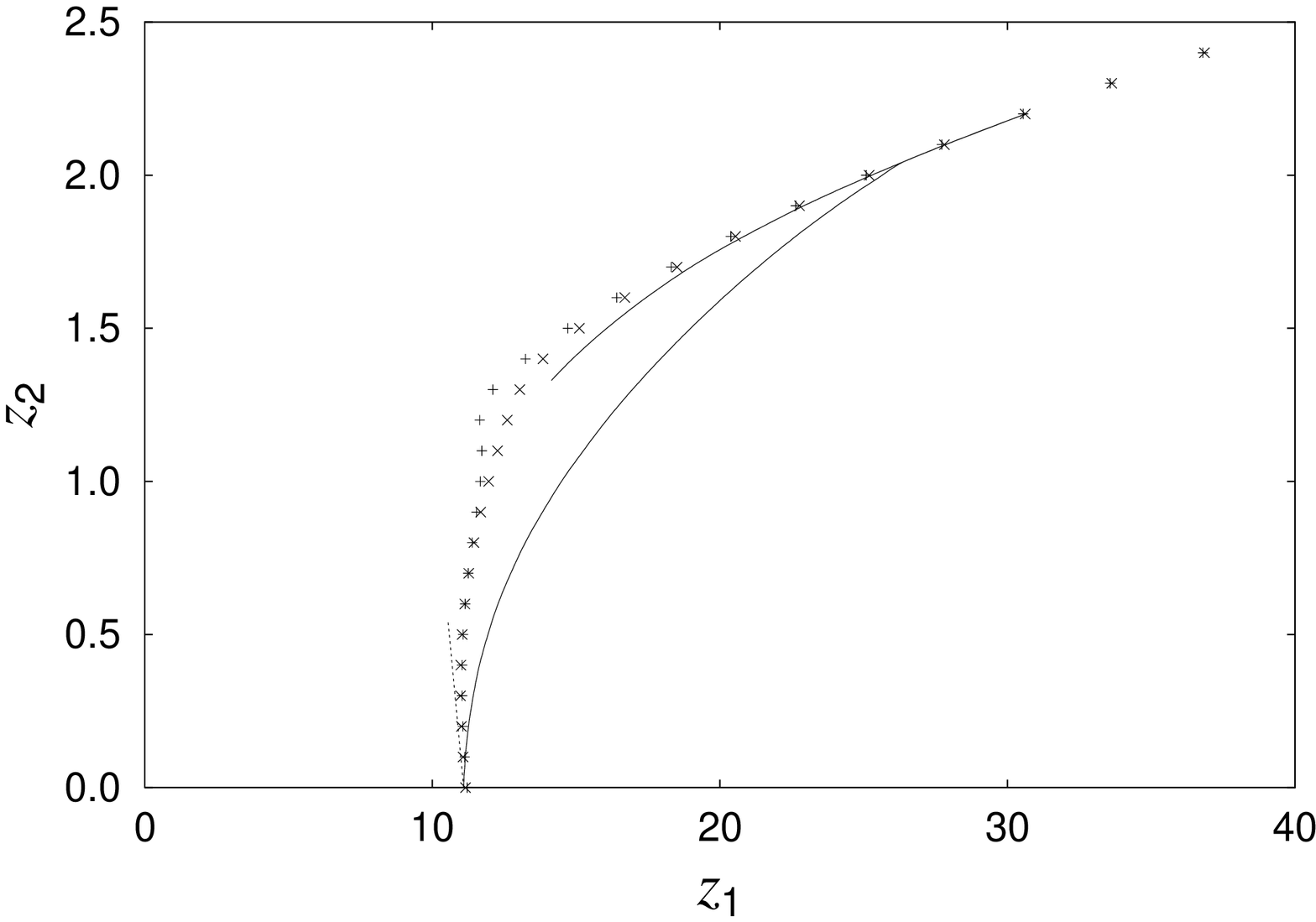}\hfil
  \caption{Locus in the $z_1$--$z_2$ plane of the minimum of the gap
    $\Delta_{\mathrm{T}}$ for $W=5$ ($+$) and $W=6$ ($\times$) and
    phase diagram of Van Duijneveldt and Lekkerkerker (solid line).
    The asymptote~(\ref{equ:hexagons:locz}) is also shown.}
  \label{fig:hexagons:minz}
\end{figure}

\begin{figure}
  \hfil\includegraphics[width=0.8 \textwidth]{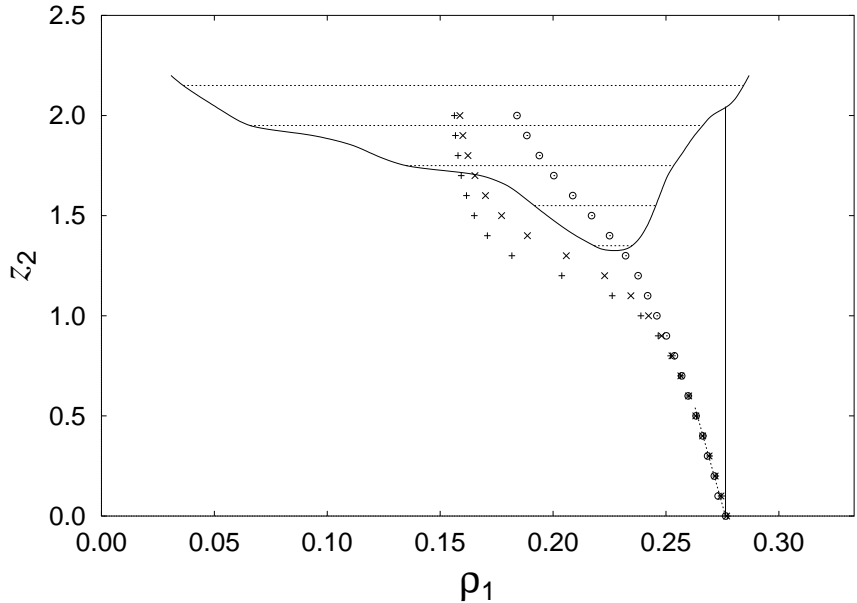}\hfil
  \caption{Locus in the $\rho_1$--$z_2$ plane of phase transition
    calculated from $W=5$ ($+$) and $W=6$ ($\times$), locus of the
    intersection of the graphs for $W=5$ and $W=6$ of $\rho_1$ versus
    $z_1$ ($\odot$), and phase diagram of Van Duijneveldt and
    Lekkerkerker (solid line).  The
    asymptote~(\ref{equ:hexagons:locrho}) is
    also shown.}
\label{fig:hexagons:minrho}
\end{figure}

The locus in the $z_1$--$z_2$ plane of the phase transition can be
estimated for example as the location of the minimum
of~$\Delta_{\mathrm{T}}$.  For fixed $z_2$ the value of $z_1$ at which
this gap takes its minimum was determined.  The results for $W=5$ and
$W=6$ are plotted in Figure~\ref{fig:hexagons:minz}.  In order to
obtain the locus in the $\rho_1$--$z_2$ plane the density of large
hexagons was computed using
\begin{displaymath}
  \rho_1 = z_1 \frac{\partial}{\partial z_1}
  \left( -\log\Lambda_0 \right).
\end{displaymath}
(It should be noted that for such small $W$ this does not seem to be
very accurate.)  Figure~\ref{fig:hexagons:minrho} shows the result.  We
observed that for fixed $z_2$ the graphs of $\rho_1$ versus $z_1$ for
different system sizes pass approximately through one point.  One could
ask whether this is the critical point, as would be the case in a
self-dual model.  The locus of the intersection of the graphs for $W=5$
and $W=6$ is shown in Figure~\ref{fig:hexagons:minrho}.  Figures
\ref{fig:hexagons:minz} and~\ref{fig:hexagons:minrho} also show the
phase diagrams given by Van Duijneveldt and
Lekkerkerker~\cite{duijneveldt:1995}.

First-order and second-order transitions are not easily distinguished
from each other by the numerical data.  In both cases
$\Delta_{\mathrm{T}}$ has a minimum, only the dependence on $L$ of the
depth of the minimum is different.  For $z_2=1.7$, the graphs of the
$L\Delta_{\mathrm{M}}$ pass approximately through one point, see
Figure~\ref{fig:hexagons:mlo}.  The $L\Delta_{\mathrm{T}}$ have a
minimum that increases slowly with~$L$ and may converge to a non-zero
value, see Figure~\ref{fig:hexagons:tlo}.  This points to a
second-order transition.  For $z_2=2.3$, the graphs of
$L\Delta_{\mathrm{M}}$ do not pass neatly through one point, see
Figure~\ref{fig:hexagons:mhi}.  The minimum of $L\Delta_{\mathrm{T}}$
decreases with~$L$ and may vanish asymptotically, see
Figure~\ref{fig:hexagons:thi}.  This points to a first-order
transition.  The behaviour of $L\Delta_{\mathrm{M}}$ and
$L\Delta_{\mathrm{T}}$ changes gradually between $z_2=1.7$ and
$z_2=2.3$.  Thus the value of $z_2$ at the tricritical point is
estimated roughly to lie between 1.7 and~2.3.

By universality the limit values of $L\Delta_{\mathrm{M}}$ and
$L\Delta_{\mathrm{T}}$ at the phase transition are
$2\pi x_{\mathrm{M}}$ and $2\pi x_{\mathrm{T}}$ respectively, with
$x_{\mathrm{M}}=2/15$ and $x_{\mathrm{T}}=4/5$ on the hard hexagon
critical line ($c=4/5$), and $x_{\mathrm{M}}=2/21$ and
$x_{\mathrm{T}}=2/7$ at the hard hexagon tricritical point ($c=6/7$),
see for instance~\cite{friedan:1984}.  On the critical line close to
the critical point one expects to find the tricritical values for small
system sizes, but the critical values for large sizes.  The limits were
also estimated from the graphs of $L\Delta_{\mathrm{M}}$ and
$L\Delta_{\mathrm{T}}$ for $z_2=0.0$ (not shown) and $z_2=1.7$.  For
$z_2=0.0$ we found $x_{\mathrm{M}} \approx 0.14$ and
$x_{\mathrm{T}} \approx 0.80$.  This is in good agreement with the
critical values $x_{\mathrm{M}}=2/15$ and $x_{\mathrm{T}}=4/5$.  For
$z_2=1.7$ we found $x_{\mathrm{M}} \approx 0.13$ and
$x_{\mathrm{T}} \approx 0.3$.  This agrees reasonably with the
tricritical values $x_{\mathrm{M}}=2/21$ and $x_{\mathrm{T}}=2/7$,
which are expected for small system size near the tricritical point.

%%%%%%%%%%%%%%%%%%%%%%%%%%%%%%%%%%%%%%%%%%%%%%%%%%%%%%%%%%%%%%%%%%%%%%%%

\section{Relation to an $A_2^{(2)}$ RSOS model}

Some properties of the large-and-small hexagon model it has in common
with an exactly solvable model.  In order to make use of the exact
solution we investigate if the two models are ever parametrically
close.  The sites of the large-and-small hexagon model can be in three
states: 0 (empty), 1 (large hexagon), or 2 (small hexagon).  For
neighbouring sites the combinations 1--1 and 1--2 are excluded.  The
same is true for the $L=7$ case of the exactly solvable $A_2^{(2)}$
restricted solid-on-solid model of
Kuniba~\cite{kuniba:1991a,kuniba:1991b}.  This is an
interaction-round-a-face model on the square lattice.  For a suitable
choice of its spectral parameter, the condition on neighbouring sites
extends to one of the diagonals of the square face.  The Boltzmann
weight of the square face then factors into weights of the composing
triangles:
\begin{displaymath}
  W\pmatrix{d & c \cr a & b \crcr} =
  W\pmatrix{d & \cr a & b \crcr} W\pmatrix{d & c \cr & b \crcr}
\end{displaymath}
and these triangle weights are invariant under rotation:
\begin{displaymath}
  W\pmatrix{c &   \cr a & b \crcr} =
  W\pmatrix{c & b \cr   & a \crcr} =
  W\pmatrix{b &   \cr c & a \crcr} =
  W\pmatrix{b & a \cr   & c \crcr} =
  W\pmatrix{a &   \cr b & c \crcr} =
  W\pmatrix{a & c \cr   & b \crcr},
\end{displaymath}
so that the model is isotropic on the triangular lattice.  The model
still has one parameter (the elliptic nome), but this solvable line
stays away from our phase diagram.  For example at the critical point
the triangle weights are
\begin{eqnarray}
  W\pmatrix{0 & \cr 0 & 0 \crcr} &=& 1, \nonumber \\
  W\pmatrix{0 & \cr 1 & 0 \crcr} &=& 4.412, \nonumber \\
  W\pmatrix{0 & \cr 2 & 0 \crcr} &=& 3.903, \nonumber \\
  W\pmatrix{0 & \cr 2 & 2 \crcr} &=& 3.129, \nonumber \\
  W\pmatrix{2 & \cr 2 & 2 \crcr} &=& 3.761, \nonumber
\end{eqnarray}
which is not of the form
\begin{eqnarray}
  W\pmatrix{0 & \cr 0 & 0 \crcr} &=& 1, \nonumber \\
  W\pmatrix{0 & \cr 1 & 0 \crcr} &=& z_1^{1/6}, \nonumber \\
  W\pmatrix{0 & \cr 2 & 0 \crcr} &=& z_2^{1/6}, \nonumber \\
  W\pmatrix{0 & \cr 2 & 2 \crcr} &=& z_2^{1/3}, \nonumber \\
  W\pmatrix{2 & \cr 2 & 2 \crcr} &=& z_2^{1/2}.  \nonumber
\end{eqnarray}
Application of the numerical transfer matrix method from
Section~\ref{sec:hexagons:transfer} to this critical model shows that
it is in the tricritical three-state Potts universality class.

%%%%%%%%%%%%%%%%%%%%%%%%%%%%%%%%%%%%%%%%%%%%%%%%%%%%%%%%%%%%%%%%%%%%%%%%

\section{Relation to the dilute three-state Potts model}
\label{sec:hexagons:potts}

\begin{figure}[!htb]
  \hfil\includegraphics[width=0.8 \textwidth]{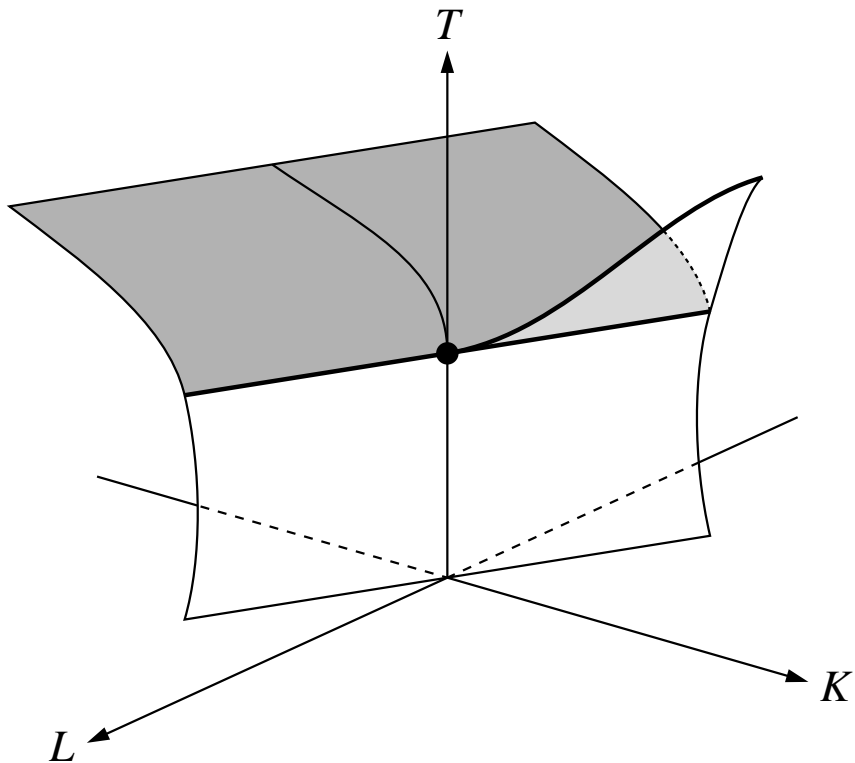}\hfil
  \caption{A qualitative picture of the phase diagram of the dilute
     three-state Potts model.  The dense coexistence region (back) and
     the dilute region (front) are separated by the three-state Potts
     critical surface (shaded) and the lower part of the first-order
     surface (not shaded).  These surfaces meet at a line of
     three-state Potts tricritical points (left) and a line of
     three-state Potts critical-end-points (right).  The upper part of
     the first-order surface (not shaded) separates a dilute and a
     dense disordered phase.  It is bounded by a line of Ising critical
     points.  The fat dot indicates the four-state Potts critical
     point.}
  \label{fig:hexagons:potts}
\end{figure}

The large-and-small hexagon model is intimately related to the dilute
three-state Potts model~\cite{berker:1978}.  Because this relation
gives insight in the phase diagram we will consider it here in more
detail.  On every site $j$ of a two-dimensional lattice with
coordination number $v$ lives a variable $s_j$ that can take the values
0, 1, 2,~3.  Of these the states $s_j>0$ take the role of local
occupancy of one of the three sub-lattices of the hard hexagon model,
and the state $s_j=0$ is neutral or vacant.  The Hamiltonian of the
dilute Potts model is
\begin{equation}
  {\cal H} = - \sum_{<j,k>} \left( \delta_{s_j,s_k} +
  K \delta_{s_j,0} \delta_{s_k,0} \right) - L \sum_j \delta_{s_j,0},
  \label{equ:hexagons:dp}
\end{equation}
where the first sum is over nearest neighbour pairs of sites.  In the
parameter space $(K,L,T)$ the model has a line of tricritical points as
well as a line of critical-end-points~\cite{berker:1978}, see
Figure~\ref{fig:hexagons:potts}.  As we will argue below, it is fairly
clear where these come together, namely in the critical point of the
four-state Potts model, $K=0$, $L=0$ and $T=T_{\mathrm{c}}$, where all
the four states are treated identically.

At $T=0$ there is a dilute phase with $s_j=0$ when $vK+2L>0$, while the
three dense, or ordered phases associated with $s_j=1,2,3$ coexist when
$vK+2L<0$.  These phases extend to non-zero temperatures so that a
first-order surface separates the dilute region from the dense
coexistence region.  This first-order surface will not remain precisely
at $vK+2L=0$ for $T>0$, but by symmetry it does include the $T$-axis,
$K=L=0$.  At high temperature the coexistence region is bounded by a
surface of three-state Potts critical points, shaded gray in
Figure~\ref{fig:hexagons:potts}, where the line tension between the
coexisting dense phases vanishes.  This critical sheet must join with
the first-order surface in a line of multicritical points, as they both
form boundaries to the coexistence region.

The nature of this multicritical line depends on the sign of $K$, as
follows.  Along the first-order sheet we can distinguish two line
tensions, namely that between two different dense phases, and that
between a
dense and the dilute phase.  When $K<0$ the interface between the
dilute and the dense phases costs less energy than that between two of
the dense phases.  However on the critical surface the line tension
between the dense phases vanishes. As a consequence all line tensions
vanish simultaneously where the critical and first-order sheets meet as
$K<0$.  The separatrix between these two types of phase transition is
thus a tricritical line.  When $K>0$ the dense--dense interface costs
less energy than the dense--dilute interface, so there remains a
positive line tension between the dilute phase and the dense phases
where the first-order sheet meets the critical surface, and the
dense--dense interfacial tension vanishes.  This results in a
critical-end-point scenario:  The three-state Potts critical sheet
terminates where it hits the first-order sheet.  The first-order sheet
extends beyond this line, separating a disordered dense phase from the
dilute phase.  Obviously, at $K=0$ the two scenarios come together,
and we conclude that the tricritical curve and the critical-end curve
as well as the critical line terminating the dilute--disordered phase
transition all meet in the four-state Potts critical point, marked as a
dot in Figure~\ref{fig:hexagons:potts}.  This qualitative description
of the phase diagram of (\ref{equ:hexagons:dp}), though not rigorous,
is the simplest possible scenario, and has been corroborated by
numerical studies~\cite{berker:1978}.

\begin{figure}
  \hfil\includegraphics[width=0.8 \textwidth]{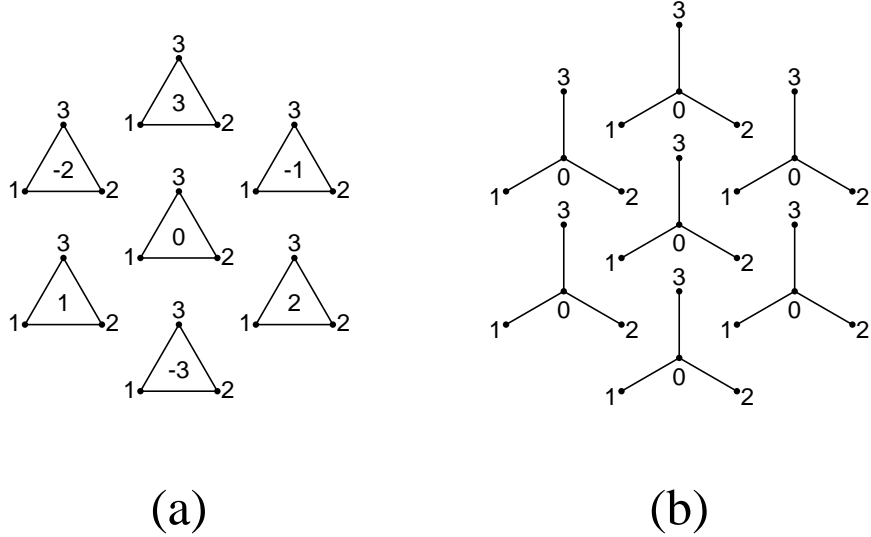}\hfil
  \caption{(a) The large-and-small hexagon model can be mapped onto a
    Potts-like model by grouping the sites into blocks of three.  The
    numbers indicate the labelling of blocks and of the sites within
    the blocks.  (b) The big hexagon model can be mapped onto a
    Potts-like model by dividing the sites into blocks of four.  The
    numbers indicate the labelling of the sites within the blocks.  The
    blocks are numbered as in (a).}
  \label{fig:hexagons:blocks}
\end{figure}

These considerations are of interest for the large-and-small hexagon
model because that can be mapped onto a model sufficiently similar to
the dilute Potts Hamiltonian (\ref{equ:hexagons:dp}) that the arguments
can be carried over.  We divide the triangular lattice into triangular
blocks of three sites each, indicated in
Figure~\ref{fig:hexagons:blocks}(a).  Each block then has three sites
which we label 1, 2, and~3.  We assign a spin variable $s_j$ to each
block, as follows.  When the site $\sigma$ in block $j$ is occupied by
a large hexagon, the spin variable takes the value $s_j=\sigma$, while
in all other cases $s_j=0$.  For convenience of notation we consider
one block variable $s_0$, in interaction with six neighbours $s_j$ with
$1 \leq |j| \leq 3 $, as shown in Figure~\ref{fig:hexagons:blocks}(a).
The blocks $j$ with $j>0$ contain two sites neighbouring the site $j$
of the central block, and the block $-j$ sits in the opposite
direction.  To give an expression for the interaction we introduce the
variables
\begin{equation}
  p_i =
  \left( \delta_{s_i,0} + \delta_{s_i,i} \right)
  \left( 1 - \delta_{s_{-j},j} \right)
  \left( 1 - \delta_{s_{-k},k} \right),
  \label{equ:hexagons:varp}
\end{equation}
where $i,j,k$ is a permutation of $1,2,3$.  Note that $p_i$ can only
take the values 0 and~1, and it signals if site $i$ of the central
block is free.  The spin states 1, 2, and 3 have weight $z_1$, but are
excluded by some configurations of the neighbouring blocks by the
factor
\begin{equation}
  \left[ 1 - \delta_{s_0,j} (1 - p_j) \right].
  \label{equ:hexagons:excl}
\end{equation}
In other words the state $s_0=j$ is not allowed when $p_j=0$.  The
weight of the spin state $s_0=0$ depends on the surrounding blocks and
is given by the expression
\begin{equation}
  (1 + z_2)^{p_1 + p_2 + p_3}.
  \label{equ:hexagons:W2hex}
\end{equation}
If this model would be precisely the dilute Potts model with
Hamiltonian (\ref{equ:hexagons:dp}) we could simply read off the value
of $K$ and its sign would conclusively decide between a tricritical
point versus a three phase point.  The interaction is of course much
more complicated than that of the dilute Potts model, but the overall
effect is that some combinations of unequal nearest neighbours are
excluded or suppressed.  As the state 0 is treated altogether different
from the states 1, 2, and 3, it is difficult to judge the sign of the
effective coupling $K$ in~(\ref{equ:hexagons:dp}).

However, this problem can be resolved because there is a model in the
universality class and with the symmetry of the four-state Potts model
which can be mapped to a very similar model.  Consider a one-species
lattice gas on the triangular lattice in which not only first
neighbours but also second neighbours (at distance $\sqrt{3}$) can not
be occupied simultaneously.  We will refer to this model as the big
hexagon model.  For large values of the fugacity $z$ this model will be
in an ordered phase in which one out of four sub-lattices is occupied
preferentially.  At low fugacity the symmetry between the sub-lattices
is unbroken.  The phase transition is known to be in the four-state
Potts universality class from the symmetry of its
Landau--Ginzburg--Wilson Hamiltonian~\cite{domany:1977,domany:1978}.
We are not aware of studies giving the critical fugacity of
this model, but we have seen numerically that it is about half the
value of the hard hexagon model.

The big hexagon model can be mapped exactly onto a Potts-like model
very similar to the model above, as expressed in
(\ref{equ:hexagons:excl}) and~(\ref{equ:hexagons:W2hex}).  Now we take
blocks of four sites as shown in Figure~\ref{fig:hexagons:blocks}(b),
one in each sub-lattice.  It is convenient to label the spins in each
block by the numbers 0, 1, 2, 3 as indicated.  When the site $j$ in a
block is occupied, the block variable takes the value~$j$.  In
addition, when none of the sites are occupied, the block variable is
taken to be~0.  Therefore the weight of the states $j>0$ is $z$ and the
weight of state 0 will again depend on the states of the neighbouring
blocks.  We again consider a block variable $s_0$ interacting with its
neighbours, which are labelled in the same way as in the previous
case.  We will use again variables $p_i$ defined by
(\ref{equ:hexagons:varp}).  The central site of the block 0 is free if
and only if $p_1=p_2=p_3=1$.  Some combinations of states of
neighbouring blocks are excluded, described by precisely the same
expression (\ref{equ:hexagons:excl}) as before.  However, also some
combinations of next-neighbouring blocks are excluded.  For example
site $j$ of block $-j$ and site $k$ of block $-k$ in
Figure~\ref{fig:hexagons:blocks}(b) are second neighbours, so the
combination $s_{-j}=j$ and $s_{-k}=k$ is excluded.  We introduce a
variable

\begin{displaymath}
  q = 1 -
  \delta_{s_{-1},1} \delta_{s_{-2},2} -
  \delta_{s_{-2},2} \delta_{s_{-3},3} -
  \delta_{s_{-3},3} \delta_{s_{-1},1} +
  2 \delta_{s_{-1},1} \delta_{s_{-2},2} \delta_{s_{-3},3}.
\end{displaymath}
Note that $q$ can only take the values 0 and~1; it signals if there are
no pairs $s_{-j}=j$ and $s_{-k}=k$.  If $s_0 \ne 0$ then $s_{-j}=j$ or
$s_{-k}=k$ is already excluded by the interaction between the
neighbouring blocks 0 and $-j$ or~$-k$.  Therefore the exclusion of the
combination $s_{-j}=j$ and $s_{-k}=k$ can be taken into account by
including a factor $q$ in the weight of block 0 in state~0.  This
weight is then given by
\begin{equation}
  q (1 + z)^{p_1 p_2 p_3}.
  \label{equ:hexagons:bigW}
\end{equation}
In this way any exclusion between sites of next-neighbouring blocks is
absorbed in the weight of state 0 of the intervening block.

This resulting model is strikingly similar to the Potts-like model
above.  The exclusion rules for pairs of neighbouring blocks are
identical and when we choose $z_1=z$, the weight of the spin states 1,
2, and 3 is the same.  In both models the weight of the state 0 depends
on the configuration of its six neighbours, via expression
(\ref{equ:hexagons:W2hex}) and (\ref{equ:hexagons:bigW}), respectively.
When we further specify $(1+z_2)^3=(1+z)$ the weights for $s_0=0$ are
equal in the case that $p_1=p_2=p_3$ and~$q=1$.  In particular they are
equal when the surrounding blocks are also in state 0, because then
$p_1=p_2=p_3=1$ and $q=1$.

It is the exclusion and suppression of configurations with unequal
neighbours which determines an effective temperature $T$ and coupling
$K$ in~(\ref{equ:hexagons:dp}).  The large-and-small hexagon model and
the big hexagon model with the parameters as set above will have the
same effective temperature $T$, as all configurations involving only
spin states $s>0$ have the same weight between the two models.  Only
when a block has $s=0$, while one or more of its neighbours have $s>0$,
the configurational weights between the two models can be different.
In all such cases the weight in the big hexagon model is smaller than
that in the large-and-small hexagon model, which is easy to see from
direct comparison of the expressions (\ref{equ:hexagons:W2hex}) and
(\ref{equ:hexagons:bigW}).  Therefore we can confidently claim that the
effective coupling $K$ is the greater in the big hexagon model, as
configurations with unequal neighbours of which one $s=0$ are more
strongly suppressed than in the large-and-small hexagon model.
However, since the big hexagon model has the symmetry of the four-state
Potts model, clearly its effective coupling $K=0$.  Therefore the
effective $K$ in the large-and-small hexagon model is necessarily
negative, which, as argued above, results in a tricritical scenario.

%%%%%%%%%%%%%%%%%%%%%%%%%%%%%%%%%%%%%%%%%%%%%%%%%%%%%%%%%%%%%%%%%%%%%%%%

\section{Discussion}

The results of our transfer matrix calculations provide evidence
against the three-phase scenario of Figure~\ref{fig:hexagons:phases}(b)
in favour of the two-phase scenario of
Figure~\ref{fig:hexagons:phases}(a).  This contradicts the earlier
findings of Van Duijneveldt and
Lekkerkerker~\cite{duijneveldt:1993,duijneveldt:1995}.  We propose the
following explanation.  Van Duijneveldt and Lekkerkerker effectively
calculate the free energy difference between the binary mixture and the
pure hard hexagons.  They then look for phases of equal pressure and
fugacities but different composition.  They do not calculate the order
parameter for the mixture.  Their method has some drawbacks.  Firstly,
it cannot detect second-order transitions, because these do not involve
a jump in the particle densities.  Secondly, it uses a polynomial fit
for the free energy difference, so that the total free energy still
seems to possess the singularity of the pure hard hexagon model.
Thirdly, whether $P$ exhibits a Van der Waals loop or not may depend
sensitively on $p(N_{\mathrm{f}}|N_1 )$.  Thus the locus of the
liquid--solid branch in their phase diagram is a spurious consequence
of the implicit assumption that the ordering transition remains at
fixed~$\rho_1$ for small values of~$z_2$.  Their qualitative conclusion
that a gas--liquid transition is present relies on quantitative
properties of the calculated phase diagram, viz.\ the locations of the
various branches.  Figure~\ref{fig:hexagons:minz} suggests that their
gas--liquid and gas--solid branch together form the true fluid--solid
line and that the critical point of their gas--liquid branch is in fact
the tricritical point.  This agrees well with the fact that Figures
\ref{fig:hexagons:minz} and~\ref{fig:hexagons:minrho} show enhanced
size dependence of the phase diagram near their gas--liquid critical
point.  However, this point is located at $z_2=1.36$ (and $z_1=22.5$),
whereas we estimate roughly $1.7<z_2<2.3$ for the tricritical point.
Being unable to present a satisfactory explanation for this
discrepancy, we stress that our data do not signal a clearly determined
locus of the tricritical point.  It should also be noted that in our
transfer matrix calculations only very small system sizes have been
considered.  Going to significantly larger systems might allow for more
definitive quantitative statements, but this requires much greater
computational resources.

Other evidence comes from the relation with the dilute three-state
Potts model.  The large-and-small hexagon model can be mapped onto a
Potts-like model.  Another model, the big hexagon model, whose phase
behaviour is known, can also be mapped onto a Potts-like model.  A
comparison of the effective temperature and coupling constants between
the large-and-small hexagon model on the one hand and the big hexagon
model on the other hand indicates that the large-and-small hexagon
model should follow the two-phase scenario.

%%%%%%%%%%%%%%%%%%%%%%%%%%%%%%%%%%%%%%%%%%%%%%%%%%%%%%%%%%%%%%%%%%%%%%%%

\subsection*{Acknowledgement}

We thank Jeroen van Duijneveldt and Henk Lekkerkerker for useful
discussions.  The former kindly provided data used in making Figures
\ref{fig:hexagons:minz} and~\ref{fig:hexagons:minrho}.  This work was
supported by the foundation Fundamenteel Onderzoek der Materie.

%%%%%%%%%%%%%%%%%%%%%%%%%%%%%%%%%%%%%%%%%%%%%%%%%%%%%%%%%%%%%%%%%%%%%%%%

\section*{Appendix}

It is instructive to follow the method of Van Duijneveldt and
Lekkerkerker using (\ref{equ:hexagons:exa}) and
(\ref{equ:hexagons:exb}) instead of Monte Carlo results.  Calculating
the pressure from (\ref{equ:hexagons:exa}) gives
\begin{equation}
  P = P^0 + \left( \rho_{\mathrm{f}} - { \mathrm{d} \rho_{\mathrm{f}}
  \over \mathrm{d} \rho_1 } \right) z_2 + o(z_2).
  \label{equ:hexagons:exc}
\end{equation}
Baxter~\cite[p.~451]{baxter:1982} lists expansions around the critical
point of several thermodynamic quantities of the pure hard hexagon
model.  Combining these expansions with (\ref{equ:hexagons:exb}) and
(\ref{equ:hexagons:exc}) yields
\begin{eqnarray}
  P &=& \left\{ P^{\mathrm{c}} + {25 (\sqrt{5}-1) \over 2
  \root{4}\of{5}} \mathop{\rm sgn} (\rho_1-\rho_1^{\mathrm{c}})
  |\rho_1-\rho_1^{\mathrm{c}}|^{3/2} + {\cal O} \left(
  (\rho_1-\rho_1^{\mathrm{c}})^2 \right) \right\} \nonumber\\
  && + \; \left\{ {125 \root{4}\of{5} (\rho_1^{\mathrm{c}})^2 \over 2
  z_1^{\mathrm{c}} } |\rho_1-\rho_1^{\mathrm{c}}|^{1/2} + {\cal O}
  (\rho_1-\rho_1^{\mathrm{c}}) \right\} z_2 \; + \; o(z_2).
\end{eqnarray}
This suggests that for small non-zero values of $z_2$ the pressure $P$
would exhibit a Van der Waals loop, so that the transition becomes
first-order as soon as $z_2$ becomes non-zero.  That this argument is
not valid can be seen by considering for example
\begin{displaymath}
  f_z(x) = (x-z)^3,
\end{displaymath}
which we view as a function of $x$, parametrically dependent on~$z$.
Expanding $f$ to first order in $z$ gives
\begin{displaymath}
  f_z(x) = x^3 - 3 x^2 z + o(z)
\end{displaymath}
and for all non-zero values of $z$ the function $x^3 - 3 x^2 z$ of $x$
is decreasing between $x=0$ and $x=2z$.  It is, however, a first order
approximant of $f_z(x)$, which for all values of $z$ is an increasing
function of~$x$.

%%%%%%%%%%%%%%%%%%%%%%%%%%%%%%%%%%%%%%%%%%%%%%%%%%%%%%%%%%%%%%%%%%%%%%%%

%%%%%%%%%%%%%%%%%%%%%%%%%%%%%%%%%%%%%%%%%%%%%%%%%%%%%%%%%%%%%%%%%%%%%%%%

\end{document}